\documentclass[reprint,prb,superscriptaddress,amsmath,amssymb,aps,floatfix]{revtex4-2}
\usepackage{graphicx}
\usepackage{bm}
\usepackage{dcolumn}
\usepackage{hyperref}
\usepackage{float}
\usepackage{siunitx}
\usepackage[caption = false]{subfig}
\usepackage{xcolor}
\usepackage{textcomp}
\usepackage{gensymb}

\begin{document}

\title{Superballistic electron flow through a point contact in a Ga[Al]As heterostructures}

\author{Lev V. Ginzburg}
\email{glev@phys.ethz.ch}
\affiliation{Department of Physics, ETH Zurich, Otto-Stern-Weg 1, 8093 Zurich, Switzerland}
\author{Carolin Gold}
\affiliation{Department of Physics, ETH Zurich, Otto-Stern-Weg 1, 8093 Zurich, Switzerland}
\author{Marc P. R\"o\"osli}
\affiliation{Department of Physics, ETH Zurich, Otto-Stern-Weg 1, 8093 Zurich, Switzerland}
\author{Christian Reichl}
\affiliation{Department of Physics, ETH Zurich, Otto-Stern-Weg 1, 8093 Zurich, Switzerland}
\author{Matthias Berl}
\affiliation{Department of Physics, ETH Zurich, Otto-Stern-Weg 1, 8093 Zurich, Switzerland}
\author{Werner Wegscheider}
\affiliation{Department of Physics, ETH Zurich, Otto-Stern-Weg 1, 8093 Zurich, Switzerland}
\author{Thomas Ihn}
\affiliation{Department of Physics, ETH Zurich, Otto-Stern-Weg 1, 8093 Zurich, Switzerland}
\author{Klaus Ensslin}
\affiliation{Department of Physics, ETH Zurich, Otto-Stern-Weg 1, 8093 Zurich, Switzerland}

\date{\today}

\begin{abstract}
    We measure electronic transport through point contacts in a high-mobility electron gas in a Ga[Al]As heterostructure at different temperatures and bulk electron densities. The conductance through all point contacts increases with increasing temperature in a temperature window around $T \sim \SI{10}{K}$ for all investigated electron densities and point contact widths. For high electron densities this conductance exceeds the fundamental ballistic limit (Sharvin limit). These observations are in agreement with a viscous electron transport model and previous experiments in graphene.
\end{abstract}

\maketitle

\section{Introduction}

Various electron transport regimes can be observed by changing the temperature and carrier density of a two-dimensional electron gas (2DEG). Two widely used models of electron transport imply either ballistic or diffusive electron flow. The ballistic transport model describes the situation where electron-electron interactions are irrelevant, and the electron mean free path related to momentum relaxation is much larger than the characteristic sample size. In turn, the diffusive model represents the non-interacting case, where momentum relaxation occurs mostly inside the system rather than at the boundaries. Beyond that, in some materials electron-electron scattering can be the dominant scattering process within a certain range of temperatures. In this case the electron-electron scattering length is much shorter than both the transport mean free path and the characteristic sample size. This transport regime is known as the regime of viscous electron flow \cite{Gurzhi63}.

The viscous regime has been widely investigated in graphene \cite{Bandurin2016, Sulpizio2019, Berdyugin2019, Ku2020}, where it is most pronounced at temperatures around $\SI{150}{K}$. It was also shown in Ga[Al]As heterostructures using the Gurzhi effect \cite{Molenkamp95, Gusev2018} or Stokes flow \cite{Gusev2020}, or based on a geometry which enables the measurement of the vicinity resistance \cite{Braem2018}. In these experiments the temperature window was centered around $\SI{10}{K}$. There are also reports of viscous transport in several other materials, including PdCoO$_2$ \cite{Moll2016}, WP$_2$ \cite{Gooth2018} and WTe$_2$ \cite{Vool2020}. One hallmark of viscous flow that has been shown in graphene \cite{KrishnaKumar2017} but not yet in Ga[Al]As is the so-called superballistic flow through a point contact (PC) \cite{Guo2016}. In this case, the conductance through the PC can be increased above the fundamental ballistic limit (Sharvin conductance). Here we demonstrate this characteristic increase of the conductance through a PC in a Ga[Al]As heterostructure in a temperature window around $\SI{10}{K}$. These results add support to the fact that viscous flow can be observed in Ga[Al]As heterostructures.

\section{Sample}

The sample is a Ga[Al]As heterostructure with a 2DEG buried $\SI{200}{n m}$ below the surface. The global patterned back-gate, roughly $\SI{1}{\micro m}$ below the 2DEG, allows us to change the electron density  from $\SI{1.5e11}{c m^{-2}}$ to  $\SI{2.7e11}{c m^{-2}}$ \cite{Berl2016}. The sample has several top-gate defined PCs in series, with a lithographic width $d$ ranging from $\SI{0.75}{\micro m}$ to $\SI{2.5}{\micro m}$. The electronic width may differ from $d$ depending on the details of the geometry and the applied gate voltages, but it remains much larger than the Fermi-wavelength in all measurements shown in this paper. For any given experiment discussed here, only a single PC was used at a time, and the remaining, unused, top-gates were grounded.

\section{Measurements and discussion}

\begin{figure*}
    \centering
    \includegraphics[width=\linewidth]{{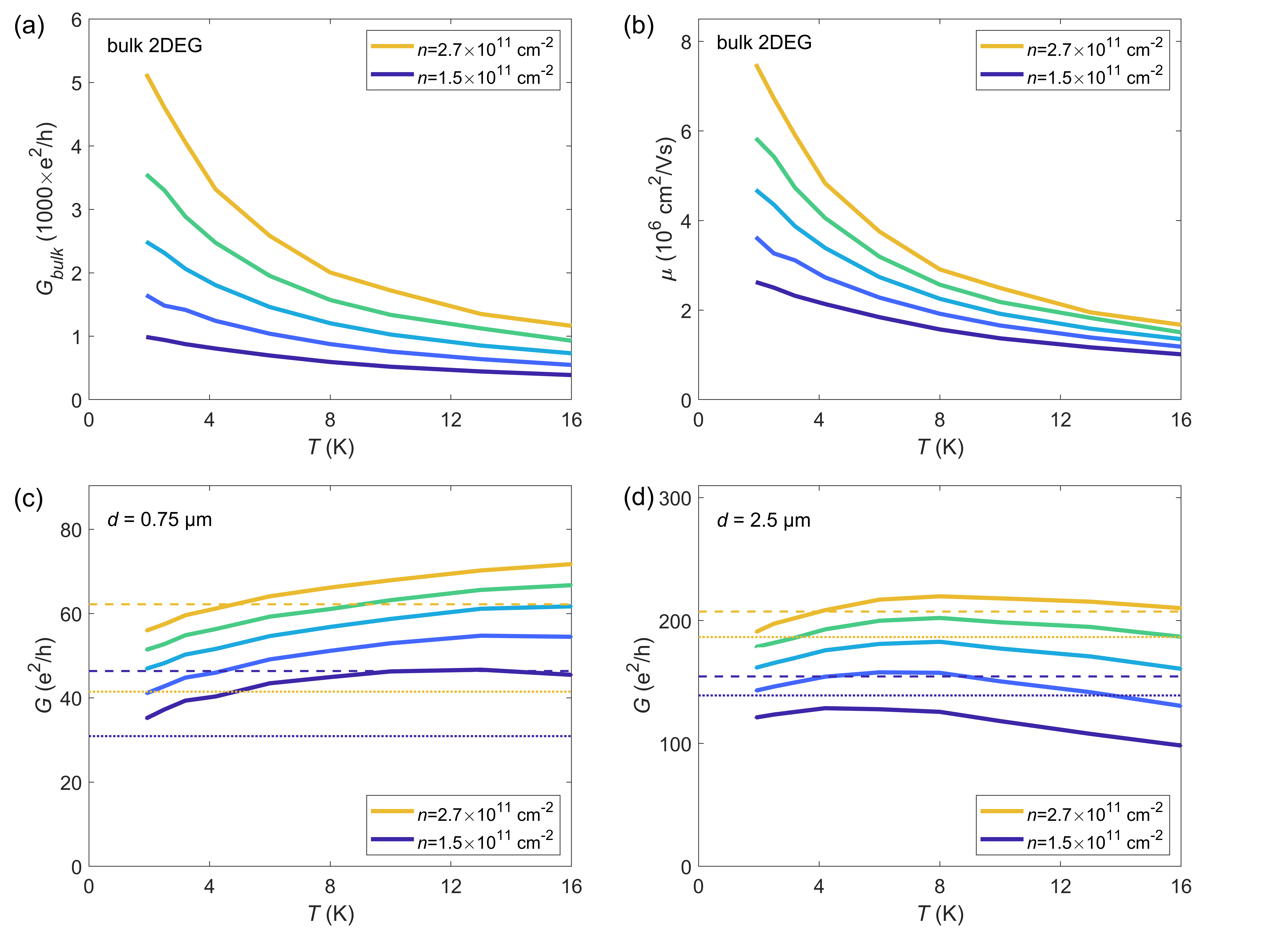}}
    \caption{(a) Bulk conductance of the sample (all top-gates are grounded) as a function of temperature for electron densities from $\SI{1.5e11}{cm^{-2}}$ to $\SI{2.7e11}{cm^{-2}}$ changed in steps of $\SI{0.3e11}{cm^{-2}}$. (b) Bulk mobility of the 2DEG. (c) Conductance $G$ of the narrowest PC ($d = \SI{0.75}{\micro m}$) and (d) conductance of the widest PC ($d = \SI{2.5}{\micro m}$) as a function of temperature; horizontal lines show the calculated Sharvin limit of PC conductance for the lithographic (dashed lines) and effective electronic (dotted lines) widths at the highest and lowest displayed charge densities, see color code.}
    \label{fig:1}
\end{figure*}

All linear conductance measurements were performed in $^4$He systems at temperatures between $\SI{1.9}{K}$ and $\SI{16}{K}$ using standard lock-in techniques at 27 Hz. The carrier densities measured using the classical Hall effect were found to be independent of temperature in the investigated range. The calibration of the measurement equipment is verified using integer quantum Hall effect. Figure~\ref{fig:1}(a) shows the conductance $G_\mathrm{bulk}$ of the 2DEG resulting from a four-terminal measurement for various carrier densities tuned by the back-gate. The conductance and with it the mobility $\mu$, see Figure~\ref{fig:1}(b), decreases with increasing temperature as a result of acoustic phonon scattering \cite{Stormer1990}.

Next, we describe the temperature dependence of the two-terminal conductance $G$ through a single PC with lithographic size $d$ for the same bulk  electron densities $n$ as in (a) and (b). Figures~\ref{fig:1}(c,d) show the results for the narrowest ($d = \SI{0.75}{\micro m}$) and widest ($d = \SI{2.5}{\micro m}$) PCs in our sample. Below a density- and width-dependent threshold temperature, the conductance $G$ increases with temperature $T$ for every set of $n$ and $d$, above it decreases. In general, the increase of the conductance with temperature is more pronounced and occurs over a larger range of temperatures for high electron density $n$ and small PC width $d$. This increase of $G$ with increasing temperature is in contrast to the decrease of the bulk conductance of the 2DEG (Figure~\ref{fig:1}(a)).

We discuss this behavior based on the geometry of the sample and the scattering lengths of electrons. At low temperatures, electron transport is ballistic and electron-electron interactions are irrelevant. The PC conductance $G$ will always be smaller than or equal to the Sharvin conductance \cite{Sharvin1965}
\begin{equation}
    G_{\mathrm{Sh}}=\frac{2e^2}{h}\times\sqrt{\frac{2nd^2}{\pi}}
\label{eq:sharvin}
\end{equation}
shown by the horizontal lines in Figures~\ref{fig:1}(c,d) for the highest ($n=\SI{2.7e11}{c m^{-2}}$) and lowest ($n=\SI{1.5e11}{c m^{-2}}$) displayed electron densities. The dashed lines show $G_{\mathrm{Sh}}$ calculated for the lithographic width of the PC. In reality, the PC width and the electron density in the PC will be slightly smaller because of side depletion and stray field effects of the gates. Therefore, we numerically calculated the effective electronic width of the PC with the software package COMSOL within the Thomas-Fermi approximation \cite{Kittel}. For the gate voltages applied in our sample, we found values which are about $\SI{250}{\nano m}$ smaller than the lithographic width. The dotted horizontal lines in Figures~\ref{fig:1}(c,d) show $G_{\mathrm{Sh}}$ for this effective width. Our observation that $G$ increases with temperature, eventually exceeding the Sharvin limit, agrees with observations in graphene \cite{KrishnaKumar2017}. Theoretical calculations \cite{Guo2016} relate this behavior to electron--electron interaction effects, which result in a hydrodynamic behavior of the electron fluid. Within this interpretation we can say that our data exhibit superballistic flow in a Ga[Al]As PC.

The decrease of the conductance with temperature above the threshold temperature can be explained in the following way. The measured two-terminal resistance comprises not only the PC resistance discussed above, but also the resistance of the 2DEG between the PC and the source/drain contacts, which is in series to the PC resistance and therefore increases it. The resistance of the bulk 2DEG can be neglected at low temperatures, but electron-phonon scattering increases the bulk resistance at higher temperatures so that the measured conductance $G$ increases more slowly with temperature or even starts decreasing (Figures~\ref{fig:1}(c,d)). This effect is expected to be more pronounced for wider PCs \cite{KrishnaKumar2017}, which is consistent with the experimental data in Figures~\ref{fig:1}(c,d).

The PC conductance in turn can be approximated as a sum of ballistic $G_\mathrm{Sh}$ and viscous $G_\mathrm{vis}$ contributions \cite{Guo2016}. Therefore, the total measured conductance $G$ can be expressed as \cite{KrishnaKumar2017}
\begin{equation}
    \frac{1}{G(T)} = \frac{1}{G_\mathrm{Sh}+G_\mathrm{vis}(T)} + \frac{1}{G_\mathrm{2DEG}(T)}.
\label{eq:sum}
\end{equation}
Here $G_\mathrm{2DEG}$ is the conductance of the bulk 2DEG between the PC and the source/drain contacts. As discussed above, $G_\mathrm{Sh}$ is independent of temperature and $G_\mathrm{vis}$ increases with temperature. The 2DEG conductance is dominated by diffusive contributions and has the same temperature dependence as the bulk conductance $G_\mathrm{bulk}$ shown above (Figure~\ref{fig:1}(a)). We did numerical calculations of $G_\mathrm{2DEG}$ with the COMSOL software package for the known geometry and measured electron mobility and density, which gave us values of $G_\mathrm{2DEG}$ down to $\SI{0.01}{\Omega^{-1}}$ at temperatures above $\SI{10}{K}$. The contribution $G_\mathrm{vis}$ was calculated from experimental data and numerical simulations. It is important to note that the result depends significantly on the exact parameters for the simulations. In order to illustrate this, we show an example of our calculations for the electron density $n=\SI{2.7e11}{c m^{-2}}$ in more detail below.

\begin{figure*}
    \centering
    \includegraphics[width=\linewidth]{{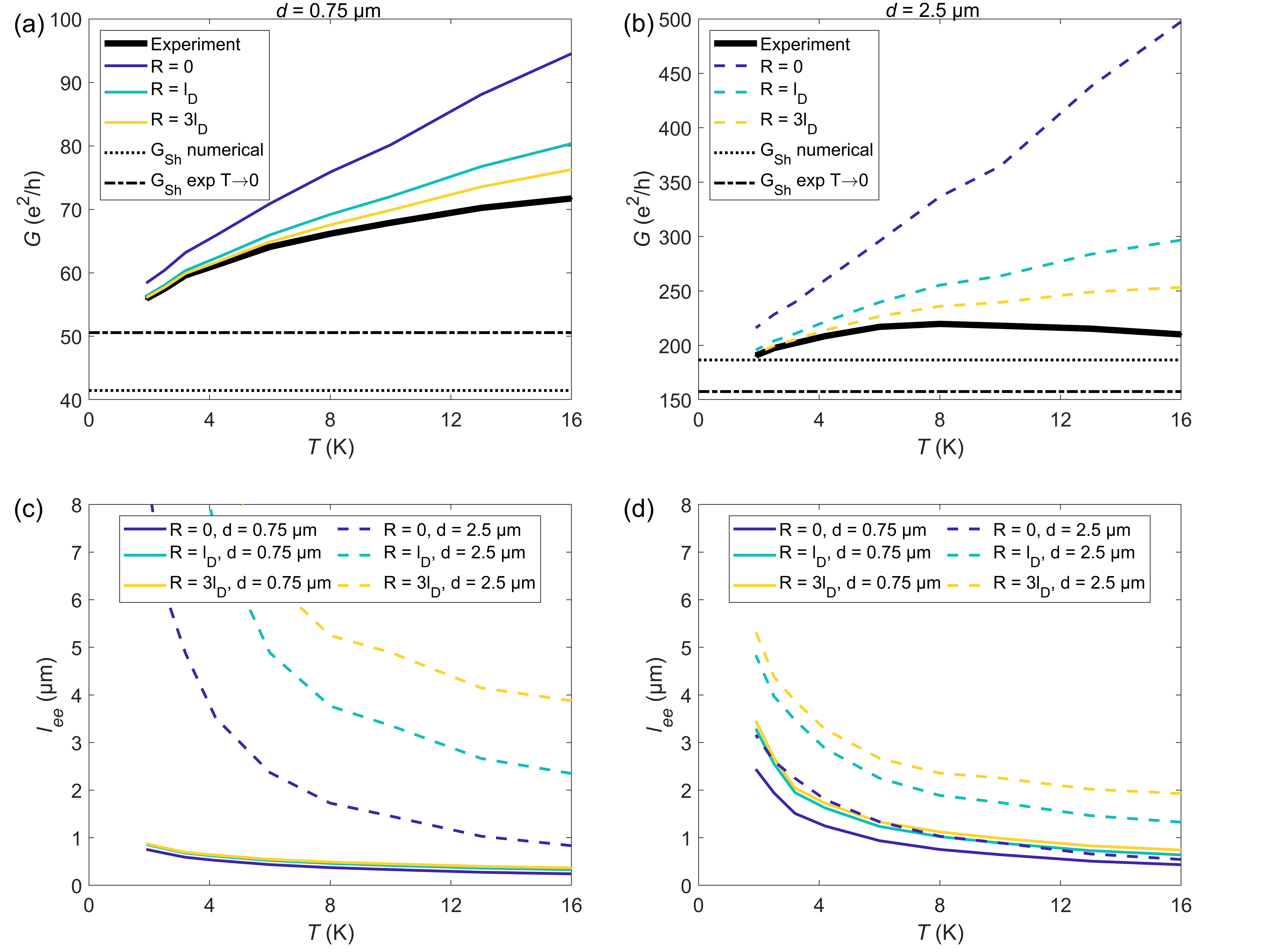}}
    \caption{Extraction of channel conductance and electron-electron scattering length from experimentally measured conductance and model simulations, all for the highest electron density $\SI{2.7e11}{cm^{-2}}$. (a), (b) Measured PC conductance as a function of temperature (thick black line) and the sum of ballistic $G_\mathrm{Sh}$ and viscous $G_\mathrm{vis}$ contributions (thin colored lines; subtracted diffusive contribution $G_\mathrm{2DEG}$ calculated numerically). Different colors correspond to different radii of the excluded circle of 2DEG $R$ around the PC in the model. (c) Estimate of $l_\mathrm{ee}$ for different $R$ and $d$. Effective width of the PCs calculated numerically. (d) Estimate of $l_\mathrm{ee}$ for different $R$ and $d$. Effective width of the PCs calculated as a low temperature limit of the experimental data.}
    \label{fig:2}
\end{figure*}

We start with the experimental curve $G$ as a function of $T$ for two PC widths (thick black line in Figure~\ref{fig:2}(a) for $d = \SI{0.75}{\micro m}$ and Figure~\ref{fig:2}(b) for $d = \SI{2.5}{\micro m}$). Next, we numerically calculate the diffusive contribution of the parts of 2DEG adjacent to the PC. However, the diffusive model is only correct on a length scale larger than the electron transport mean free path $l_\mathrm{D}$. Therefore, we choose a circle of radius $R$ around the PC; within our model we consider only the diffusive contribution of the 2DEG outside this circle. The choice of the radius $R$ should be comparable to $l_\mathrm{D}$ and is arbitrary to some extent, but important for the estimate of $l_\mathrm{ee}$. In Figures~\ref{fig:2}(a,b) in blue, green and yellow colors (dark gray, gray and light gray for grayscale version) we show $(1/G-1/G_\mathrm{2DEG})^{-1}$ for $R = 0$, $l_\mathrm{D}$ and $3l_\mathrm{D}$ correspondingly. Here $G$ is measured data, $G_\mathrm{2DEG}$ is calculated numerically. The conductivity of the 2DEG and the transport mean free path are determined from the known mobility and electron density (see Figure~\ref{fig:1}(b)).

According to Eq.~(2), the resulting curves should represent the sum of ballistic and viscous contributions $G_\mathrm{Sh} + G_\mathrm{vis}$. For all six curves (two $d$ values and three $R$ values), the conductance increases monotonically, which is in agreement with theoretical predictions \cite{Guo2016}. In the next step we subtract the ballistic part of the conductance $G_\mathrm{Sh}$ (dotted black lines in Figures~\ref{fig:2}(a,b)) determined for the effective width of the PC $d_\mathrm{eff}$ numerically calculated with the electrostatic COMSOL model ($d_\mathrm{eff} = \SI{0.50}{\micro m}$ and $d_\mathrm{eff} = \SI{2.25}{\micro m}$ for the narrowest and widest PC respectively). The resulting $G_\mathrm{vis}$ is then used to calculate the electron-electron mean free path $l_\mathrm{ee}$ according to \cite{Guo2016}
\begin{equation}
    l_\mathrm{ee} = \frac{e^2d^2_\mathrm{eff}}{8\hbar}\sqrt{\frac{\pi n}{2}}\frac{1}{G_\mathrm{vis}}.
\label{eq:lee}
\end{equation}
The resulting $l_\mathrm{ee}$ for all six cases is shown in Figure~\ref{fig:2}(c). These curves should be independent of $d$, however one can see that the values for different PC widths are not close to each other in this calculation.

In order to improve the model, we use the estimate of the zero temperature limit of the $G_\mathrm{Sh} + G_\mathrm{vis}$ (dashed-dotted black lines in Figures~\ref{fig:2} (a,b)), obtained by extrapolating the experimental conductance in Figs.~\ref{fig:2}(a,b) to zero temperature, instead of the results of numerical calculations as the value of $G_\mathrm{Sh}$ and determined $d_\mathrm{eff}$ from it ($d_\mathrm{eff} = \SI{0.61}{\micro m}$ and $d_\mathrm{eff} = \SI{1.90}{\micro m}$ for the narrowest and widest PC respectively). Figure~\ref{fig:2}(d) shows the $l_\mathrm{ee}$ obtained in this way. While the resulting curves are closer to each other than in Figure~\ref{fig:2}(c), showing an improvement in the model, they still differ markedly.

We conclude from this analysis that the estimate of the electron-electron mean free path depends considerably on the model parameters $R$ and $d_\mathrm{eff}$, which we do not know precisely enough. A precise calculation of $l_\mathrm{ee}$ is beyond the scope of our model. Nevertheless, we observe the increase of $G_\mathrm{vis}$ with increasing temperature at all temperatures and we can make an order of magnitude estimate for the electron-electron scattering length $l_\mathrm{ee}$, which changes from $\sim \SI{5}{\micro m}$ at low temperatures $T=\SI{2}{K}$ to $\sim \SI{1}{\micro m}$ at high temperatures $T=\SI{15}{K}$. This result is comparable to theoretical estimates of this value \cite{BraemThesis}.

Superballistic flow has been observed before in graphene \cite{KrishnaKumar2017}. It turns out that the overall phenomenology in GaAs and graphene is very similar from an experimental point of view. The difference mostly lies in the characteristic temperature window (10 K for GaAs versus 150 K for graphene) where viscous flow dominates the PC conductance. This can be attributed to the different electron-phonon coupling and the characteristic phonon frequencies in these two materials.

\section{Conclusions}

We performed measurements of the conductance through point contacts in a Ga[Al]As 2DEG for different temperatures, bulk electron densities and point contact widths. We consistently observe an increase of the conductance with increasing temperature within a certain temperature interval around $T \sim \SI{10}{K}$ for all available values of the electron density and the point contact width. For high electron densities the point contact conductance exceeds the fundamental ballistic (Sharvin) limit. We interpret these observations within the viscous electron transport model as superballistic flow through the point contact.

\section{Acknowledgments}

The authors acknowledge financial support from Eidgenössische Technische Hochschule Zürich (ETH Zurich) and the Swiss National Science Foundation via National Center of Competence in Research Quantum Science and Technology (NCCR QSIT)

\bibliographystyle{apsrev4-1}
\bibliography{main}

\end{document}